\documentclass[aps,prl,twocolumn,floats, showpacs,superscriptaddress,groupedaddress, 10pt]{revtex4-1}  
\usepackage{amsfonts}
\usepackage{amsmath}
\usepackage{pspicture}
\usepackage{epsfig}
\usepackage{calc}
\usepackage{times}
\usepackage{color}
\usepackage{subfigure}
\usepackage{graphicx}
\usepackage{enumerate}
\usepackage{amssymb}
\usepackage{amsthm}
\newcommand{\appref}[1]{\hyperref[#1]{{Appendix~\ref*{#1}}}}
\newcommand{\cancel}[1]{}
\newcommand*{\cC}{\mathcal{C}}
\newcommand*{\cE}{\mathcal{E}}

\newcommand*{\cH}{\mathcal{H}}

\newcommand*{\cN}{\mathcal{N}}

\newcommand*{\cR}{\mathcal{R}}

\newcommand*{\cZ}{\mathcal{Z}}

\newcommand{\bc}{\begin{center}}
\newcommand{\ec}{\end{center}}

\newcommand{\id}{\mathbb{I}}
\newcommand{\idchannel}{\mathsf{id}}

\newcommand{\Tr}{\mathop{\mathrm{tr}}\nolimits}

\usepackage{amsfonts}

\def\id{\mathbb{I}}

\def\01{\{0,1\}}

\newcommand{\ket}[1]{|#1\rangle}
\newcommand{\bra}[1]{\langle#1|}

\newcommand{\ketbra}[2]{|#1\rangle\langle#2|}

\newcommand{\comment}[1]{}

\renewcommand*{\kappa}{k}

\newcommand{\tmop}[1]{\ensuremath{\operatorname{#1}}}
\newcommand{\assign}{:=}

\definecolor{jens}{rgb}{0,.8,.5}

 \newcommand{\myappendix}[1]{#1}   

\usepackage{color}

\begin{document}

\title{Towards holography via quantum source-channel codes}
\author{Fernando \surname{Pastawski}, Jens \surname{Eisert} and Henrik \surname{Wilming}}
\affiliation{Dahlem Center for Complex Quantum Systems, Freie Universit{\"a}t Berlin, 14195 Berlin, Germany
}
\date{\today}

\begin{abstract}
While originally motivated by quantum computation, quantum error correction (QEC) is currently providing valuable  insights into many-body quantum physics such as topological phases of matter.
Furthermore, mounting evidence originating from holography research (AdS/CFT), indicates that QEC should also be pertinent for conformal field theories.
With this motivation in mind, we introduce quantum source-channel codes, which combine features of lossy-compression and approximate quantum error correction, both of which are predicted in holography.
Through a recent construction for approximate recovery maps, we derive guarantees on its erasure decoding performance from calculations of an entropic quantity called conditional mutual information.
  As an example,
  we consider Gibbs states of the transverse field Ising model at criticality and provide evidence that they exhibit non-trivial protection from local erasure.
  This gives rise to the first concrete interpretation of a bona fide conformal field theory as a quantum error correcting code.
 We argue that quantum source-channel codes are of independent interest beyond holography.
 %

\end{abstract}
\pacs{}
\maketitle

Driven by the prospects of building a large scale quantum computer, the field of quantum information \cite{Nielsen2000} has flourished, and within it \textit{quantum error correction} (QEC).
QEC is the framework for quantum \textit{channel-coding}, and allows us to understand how noisy degrees of freedom (DoFs) may be used to implement a smaller number of reliable DoFs \cite{Preskill1997}.
In contrast, \textit{compression} (source-coding) describes how statistical redundancies may be exploited to reduce the naive number of required resources.
Seeking separation of concerns, these two techniques are conventionally deal with separately, particularly in the classical regime where vast volumes of data justify the use of Shannon's source-channel separation theorem \cite{Shannon1948}.

In this work, we depart from the customary approach of separating source and channel coding. We justify this from two perspectives.
From a coding theory perspective, joint source-channel coding provides improved performance when the source is not an asymptotic i.i.d.\ distribution \cite{Modestino1981}.
Our second motivation comes from fundamental physics; more specifically, from quantum-gravity.

In the holographic approach \cite{Susskind1995, Hooft1993, Maldacena1998, witten1998anti}, quantum gravity on a $d+1$ dimensional asymptotically Anti-de Sitter (AdS) background is understood to be isomorphic to certain conformal field theories (CFTs) in $d$ dimensions.
For example, the vacuum state in the CFT corresponds to ("is dual to") the AdS vacuum, while thermal states in the CFT correspond to a static black holes in the center of AdS \cite{Maldacena2003,VanRaamsdonk2017}.
It has been argued that this isomorphism, dubbed AdS/CFT correspondence, shows both features of quantum error correction \cite{Almheiri2015} and data compression \cite{Czech2015}.

It is difficult to reconcile how error correction features can show up in a single state of a CFT such as the vacuum or a full rank thermal state from the perspective of traditional Quantum Error Correction Codes (QECCs).
The main difficulty lies in the absence of a proper subspace of full Hilbert space, where logical information can be deposited.
We require a distinct framework and introduce the name \textit{source-channel codes}, to emphasize that the task being achieved is different to that of traditional QECCs and closely follows earlier work by Barnum and Knill \cite{Barnum2002} instead.
Through an example, we show that error-correcting properties of CFTs can be made explicit in the framework of \emph{quantum source-channel codes}.
Our analysis confirms predictions from holography, even beyond its expected regime of validity.

Our presentation is structured as follows.
We first review some basics about source coding and channel coding from a purely information theoretic perspective.
We then discuss source-channel coding and show, using recent results from quantum information theory, how recovery guarantees can be given by calculating entropies.
Finally, guided by the QECC interpretation of AdS/CFT \cite{Almheiri2015,Pastawski2015,Hayden2016,Harlow2016,Pastawski2017} we consider thermal states of the transverse field Ising CFT as an example of quantum source-channel coding.
This is supported by numerical evidence of non-trivial error correction guarantees with respect to local erasures.

{\bf Source and channel coding.}
In traditional \textit{source coding}, statistical properties of an input source distribution $X$ are identified and exploited to reduce the message size.
The standard procedure is to discard atypical inputs.
Shannon's source coding theorem \cite{Shannon1948} considers the asymptotic i.i.d. setting (i.e. the source consists of a large number $n$ of messages sampled over the same distribution $X$).
It states that $\lceil n H(X) \rceil$ bits are sufficient to support most of this product distribution (compression fails with vanishing probability), where $H(X)$ denotes the Shannon entropy of $X$;
Schumacher compression \cite{Schumacher1995} achieves this in the quantum setting.
However, if the source space is larger than the output space, compression will necessarily be lossy. In particular, compression will be either trivial or lossy for any finite $n$.

\begin{figure}[t!]
	\begin{center}
		\includegraphics[width=.7\columnwidth]{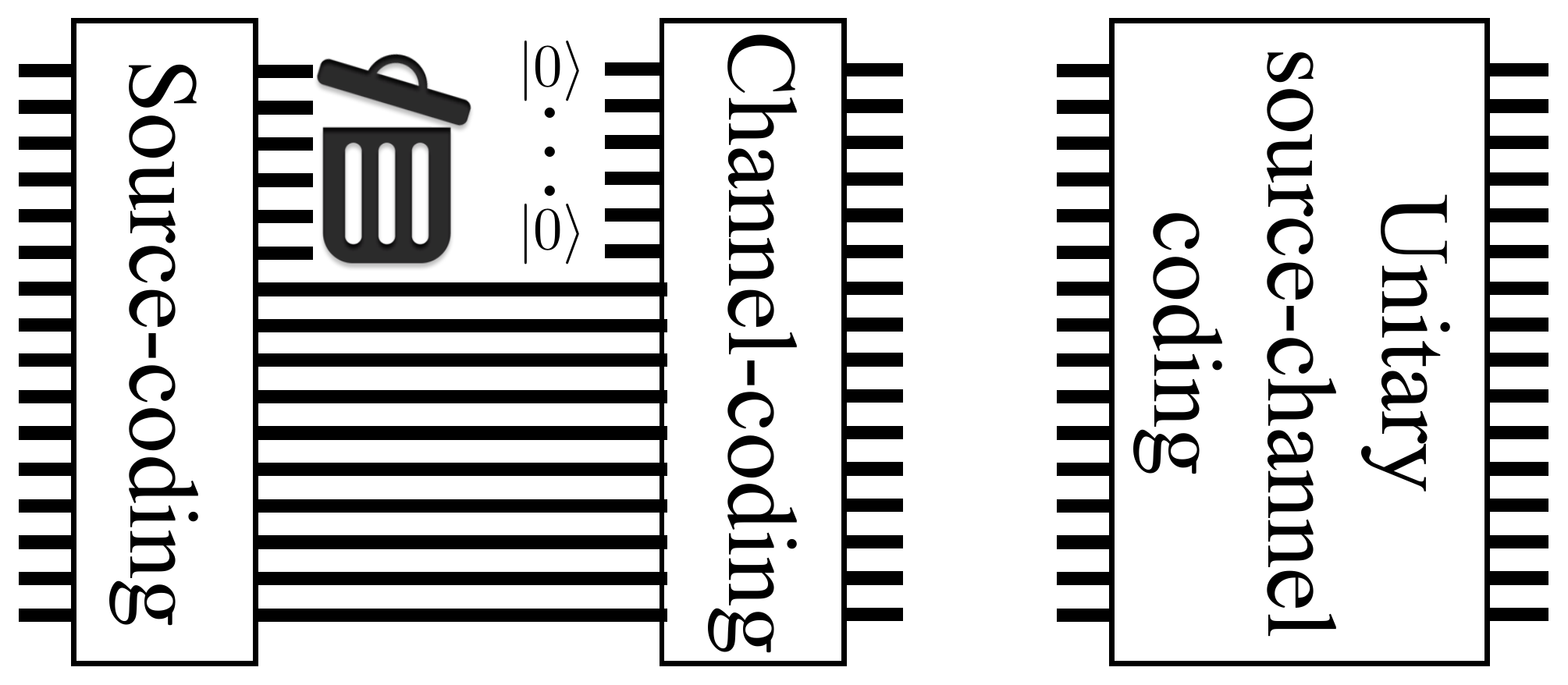}
		\caption{
			For a density matrix $\rho$, Schumacher compression may be performed by a unitary which concentrates the information of $\rho^{n}$ on the last $k\approxeq S(\rho)n$ qubits.
			As a result of the compression unitary, we may think of the remaining $n-k$ qubits as being set to zero with high probability
			and discarding them leads to minimal loss.
			Indeed, the compression can also be interpreted as an algorithmic cooling unitary \cite{Schulman1999} in which the discarded qubits approximate a known pure state.
			Precisely how small a loss will depend on $\rho$ and the choice of $k$.
			Conversely, traditional channel coding for a $[[n,k,d]]$ QECC may be performed by a unitary which takes as input $k$ data qubits and $n-k$ ancillary qubits which are initially set to $\ket{0}$.
			We may view unitary source-channel coding as the result of matching parameters and skipping the intermediate reinitialization of ancillas.
		}\label{fig:Comparison}
	\end{center}
\end{figure}

QECCs are the paragon example of (quantum) \textit{channel coding}.
For concreteness, we focus on qubit codes parametrized by $[[n,k,d]]$.
An encoding is an isometric map from $k$ qubits to $n\geq k$ qubits.
In particular, the \textit{code subspace} is a $2^k$ dimensional subspace of the $2^n$ dimensional physical Hilbert space.
For us, it will be important that such an isometry may be viewed as a unitary from $k$ logical qubits and $n-k$ ancillas onto the physical Hilbert space (see Fig. \ref{fig:Comparison} left).
In certain settings, such as when the noise channel affects no more than $d-1$ identified qubits or $(d-1)/2$ unspecified qubits, the code space allows for perfect recovery.
In general however, if one aims for \textit{approximate recovery} from the outset, it is possible to design codes which deal with a much broader set of noise channels;
this is the approach taken by \textit{approximate quantum error-correcting codes} \cite{Crepeau2005}.

Source-channel separation theorems state that it is possible to separate the problem of source coding and channel coding without any sacrifice in the achievable communication rate \cite{Shannon1948}.
While these results generalize to the quantum setting (see Ref.\ \cite{Datta2013} and references therein) they hold only in an asymptotic i.i.d. limit, assuming both the source and channel are memoryless.
In contrast, for the single shot setting, which is relevant for practical communication problems with correlated data, joint source-channel coding has been found to be advantageous \cite{Modestino1981}.
In particular, we consider the source Hilbert space to be of equal dimension as the target Hilbert space such that the combined source-channel encoding is a unitary $U$ and lossless in the case of no erasure.
This choice is also motivated by the holographic duality, which we seek to interpret as a unitary isomorphism.

We interpret a decomposition of a density matrix
$\sigma = \int_\phi \ketbra{\phi}{\phi} \mu(\phi) d\phi$
into a convex combination of pure states as specifying a \textit{source distribution}.
We will refer to the probability measure $\mu$ as a \emph{resolution of $\sigma$  into an ensemble} of pure states 
, where the von~Neumann entropy $S(\sigma)$ upper bounds the amount of classical information conveyed \cite{Holevo1973}.
We defer the difficult problem of identifying an adequate source-channel encoding unitary $U$ for a given source distribution and noise model to future work.
Our analysis focuses on the erasure recovery of an encoded distribution $\rho$ related to $\sigma$ by $\rho=U \sigma U^\dagger$.

In our approach, choosing a maximally mixed density matrix $\rho$ within a given code subspace will correspond to considering average fidelity in traditional subspace encoding QECCs (we elaborate on this in appendix A).
However, a density matrix $\rho$ with generic spectrum, (possibly supported over the full Hilbert-space) will correspond to a non-trivial prior $\sigma$ on the distribution of input states.
In this sense, our approach combines source coding with channel coding.

{\bf Fidelities.} We now review fidelity measures and their relations used to benchmark the performance of source-channel codes.
Given a noise channel $\cN$ and the recovery channel $\mathcal{R}$, one may calculate the fidelity with which a single pure state $\ket{\psi} \in {\mathcal H}_{BC}$ is recovered as
\begin{equation}\label{eq:stateFidelity}
F^2 ( \ket{\psi},\cE)
\assign
\bra{\psi} \cE ( \ketbra{\psi}{\psi}) \ket{\psi},
\end{equation}
where $\cE:=\mathcal{R}\circ \cN$ combines noise and recovery.
Given a distribution $\mu$ of pure states, the \textit{weighted average fidelity}
\begin{align}\label{eq:AvGFidelity}
\bar{F}(\mu, \cE)
:=
\int_{\psi} \mu(\psi) \bra{\psi}\cE ( \ketbra{\psi}{\psi}) \ket{\psi} d \psi,
\end{align}
 quantifies the corresponding weighted average.
To obtain bounds on the average fidelity, we make use of the \emph{Schumacher entanglement fidelity} proposed in Ref.\ \cite{Schumacher1996}.
For any density matrix $\rho= \int_\psi \ketbra{\psi}{\psi} \mu(\psi) d\psi$, it is defined as
\begin{align}\label{eq:WeightedEntanglementFidelity}
F_e(\rho, \cE) \assign F^2(\ket{\Psi_\rho},\idchannel_A \otimes \cE),
\end{align}
where $\ket{\Psi_\rho}$ is an arbitrary purification of $\rho$, i.e., a pure state on an enlarged Hilbert-space $\mathcal{H}_A\otimes \mathcal{H}_{BC}$ such that $\mathrm{Tr}_A(|\Psi_\rho \rangle\langle\Psi_\rho|)=\rho$.
If $\rho$ is a thermal state of a Hamiltonian with energies $E_j$, the \emph{thermofield double state} \cite{Maldacena2003}
\begin{equation}
\mid \Psi_{\tmop{TFD}} \rangle \assign \frac{1}{\sqrt{\mathcal{Z}}} \sum_j
e^{- \beta E_j/ 2} \mid j \rangle_A \mid j \rangle_{BC}
\end{equation}
with $\cZ := \sum_j e^{-\beta E_j}$, provides an explicit choice for such a purification.
However, definition \eqref{eq:WeightedEntanglementFidelity} is independent of the purification used on the RHS.

Schumacher~\cite{Schumacher1996} proved that the average fidelity $\bar{F}$ with respect to any resolution $\mu$ of $\rho$ is lower bounded by the
entanglement fidelity
\begin{align}\label{eq:EntanglementEnsembleBound}
F_e(\rho, \cE) \leq \bar{F}(\mu, \cE).
\end{align}
It is thus sufficient to focus on the entanglement fidelity $F_e(\rho,\cE)$ of a density matrix $\rho$ in order to provide a lower bound on the weighted average fidelity \eqref{eq:AvGFidelity} for any ensemble $\mu$ of states generating the statistics of $\rho$.
Our next task will be to guarantee the existence of a recovery map $\cR$ such that $\cE := \cR \circ \cN$ realizes a high entanglement fidelity for a given noise map $\cN$.

{\bf Markov condition and recovery.}
If we know on which subsystem $C$ the noise channel $\cN$ acts non-trivially,
we can invoke powerful prescriptions \cite{PETZ1988, Fawzi2015, Sutter2016} to specify a recovery map $\cR$.
Given a tensor product Hilbert space $\cH_A \otimes \cH_B \otimes \cH_C$, and a noise channel $\cN$ acting only on subsystem $C$, there exists a recovery map $\cR$ such that
\begin{align}\label{eq:MarkovRecovery}
I(A:C|B) \geq -2 \log F(\rho_{ABC}, \idchannel_A\otimes \cR),
\end{align}
where $I(A:C|B) := S_{AB}+ S_{BC} - S_{ABC} -S_B$ is
the conditional mutual information (CMI) of $A:C$ conditioned on $B$.
Moreover, the recovery map $\cR$ will act exclusively on $BC$, attempting to reconstruct $C$ from $B$ without access to $A$.
Intuitively, the recovery map assumes the worst possible scenario for subsystem $C$ and ignore its content altogether (i.e. $\cR = \cR\circ \cN \equiv \cE$).
For $I(A:C|B) \ll 1$, we can use eq.\ $\eqref{eq:MarkovRecovery}$ to derive
\begin{align}\label{eq:MarkovBound}
1- I(A:C|B) \leq  F^2(\rho_{ABC}, \idchannel_A\otimes \cR).
\end{align}
Furthermore, for pure $\rho_{ABC}$, the CMI simplifies to
\begin{align}\label{eq:MarkovQuantitySimplified}
I(A:C|B) = S_C + S_{BC} -S_B,
\end{align}
where we have chosen an expression without reference to $A$.
Interpreting $\cH=\cH_B\otimes \cH_C$ as the physical Hilbert space and $\cH_A$ as a fictitious ancilla space purifying $\rho_{BC}$.
This corresponds to the problem of global recovery in which $B$ includes all physical information not affected by noise $\cN$.
Assuming $\rho_{ABC} := \ketbra{\Psi_\rho}{\Psi_\rho}$ to be a pure state we recognize the entanglement fidelity \eqref{eq:WeightedEntanglementFidelity} in the RHS of eq.\ \eqref{eq:MarkovBound}.
We then have our final result
\begin{align}\label{eq:Bound}
1- S_C + S_{BC} -S_B \leq F_e(\rho, \cE) \leq \bar{F}(\mu, \cE).
\end{align}

While the general prescription \cite{Sutter2016} for $\cR$ depends on $\rho_{ABC}$, given $\rho_{BC}$ it is independent of its purification on $ABC$.
To summarize, given a state $\rho_{BC}$ with a resolution $\mu$ and a noise map $\cN$ acting exclusively on subsystem $C$, it is possible to specify a fixed recovery map $\cR$ which recovers the state with entanglement (weighted average) fidelity above $1-\epsilon$ provided $S_C + S_{BC} - S_B \leq \epsilon$.

{\bf Numerical evaluation on a CFT.}
In order to accentuate the advantage of joint source-channel coding we focus on a family of source distributions deviating from the i.i.d. property.
We observe that the two ``natural'' separations of a CFT thermal distributions into degrees of freedom are far from i.i.d.
A real space blocking of CFT degrees of freedom could lead to identical blocks which are however not independent;
CFTs present correlations at all scales.
Traditional, i.i.d. source coding would not take advantage of the correlations, leading to a significant overestimate in the total source entropy unless the block length is much larger than the CFT thermal length.
A momentum space separation of CFT degrees of freedom would lead to independent degrees of freedom for a free CFT, which are however far from being identical to one another.
This suggests that CFT thermal distributions might be good candidates for the application of source-channel coding.

We now discuss a concrete CFT model and the corresponding numerical results and later comment on the geometric interpretation suggested by holography.
The \emph{critical transverse field Ising model} on a ring
\begin{align}\label{eq:TFIsing}
H_{TF} = \sum_{j=1}^n \left( - \sigma_j^x\sigma_{j+1}^x - \sigma_{j}^z \right).
\end{align}
is arguably the simplest spin Hamiltonian approximating a CFT.
The parity operator $ P :=\bigotimes_{j=1}^n \sigma_j^z$ is a conserved quantity for $H_{TF}$ and $P_{\text{even}} := (\id+P)/2$ and $P_{\text{odd}} := (\id-P)/2$ are the corresponding projectors.
Within each parity sector, $H_{TF}$ is equivalent to a free fermion Hamiltonian
\begin{align}\label{eq:MajoranaFermionHamiltonian}
H_{\text{maj}}^{(\text{odd,even})} := \sum_{j=1}^{2n-1} i w_j w_{j+1} \pm i w_{2n}w_{1},
\end{align}
(also restricted to a specific fermionic parity sector~\cite{Pfeuty1970})
where we have introduced Majorana fermion operators $w_j$ satisfying the commutation relations $\{\omega_i,\omega_j\}=\delta_{ij}\id$.
We will consider the thermal state at inverse temperature $\beta>0$, restricted to the even parity sector
\begin{align}\label{eq:EvenThermal}
  \rho_{BC} = \rho^{(\beta)}_\text{even} := \frac{P_\text{even} e^{-\beta H_{TF}} }{\Tr{[P_\text{even} e^{-\beta H_{TF}}]}}.
\end{align}
Since the original thermal state is block diagonal with respect to $P$, we are simply normalizing the even block.

We justify the parity projection of eq.\ \eqref{eq:EvenThermal} from two perspectives.
From a physical viewpoint, $P$ distinguishes fermionic parity superselection sectors in the fermionic interpretation \eqref{eq:MajoranaFermionHamiltonian} of $H_{TF}$.
A manifestation of the fermionic Hilbert space not being of product form is that the Fermionic parity can not be changed by a \textit{physical} noise process.
However, the qubit erasure channel does not enforce this.
The global parity projection $P_{\text{even}}$ of \eqref{eq:EvenThermal} compensates for this at the level of the state in a relatively simple way.
As we will see, the parity projection is also more favorable in terms of its coding properties (particularly when we take $\beta \propto n$ as we will).
While the parity projected state $\rho^{(\beta)}_\text{even}$ seems to lead to a quadratic scaling $O(n^{-2})$ of the CMI of eq.\ \eqref{eq:MarkovQuantitySimplified}, the full thermal state results in a ``trivial'' $O(1/n)$ scaling \cite{Trivial}.

The price we pay for including the parity projection is that, contrary to the full thermal state of \eqref{eq:MajoranaFermionHamiltonian}, $\rho^{(\beta)}_{\text{even}}$ is not a fermionic Gaussian state
(see appendix B\ref{sec:GaussianStates} for a comparison to Gaussian states).
This forces us to use exact diagonalization for the numerical evaluation of the CMI \eqref{eq:MarkovQuantitySimplified}
which has a computational cost scaling exponentially in $n$ instead of polynomially.
In turn, this hinders us from considering $n > 13$ without dedicating numerical resources beyond the scope of this work.

The Hamiltonian $H_{TF}$ (respectively $H_{\text{maj}}$), has a linear dispersion relation and a gap scaling as $1/n$.
In order to maintain the global entropy  $S_{BC} = S(\rho^{(\beta)}_\text{even})$ roughly independent of the system size $n$,
it will also be necessary to scale $\beta \propto  n$
(i.e. we scale the temperature $1/\beta$ linearly with the Hamiltonian gap).
Note that in the interpretation as quantum source-channel codes, $n$, the entropy $S_{BC}$ and $|C|$ respectively play roles analog to $n$, $k$ and $d$ in traditional $[[n,k,d]]$ QECCs, where to each $d$ we may associate a guaranteed performance in terms of average fidelity (or lack thereof).
In appendix C\ref{sec:Largek} we discuss the possibility of encoding more logical information by scaling $\beta$ sub-linearly with $n$.

\begin{figure}[t]
	\begin{center}
		\includegraphics[width=0.851\columnwidth]{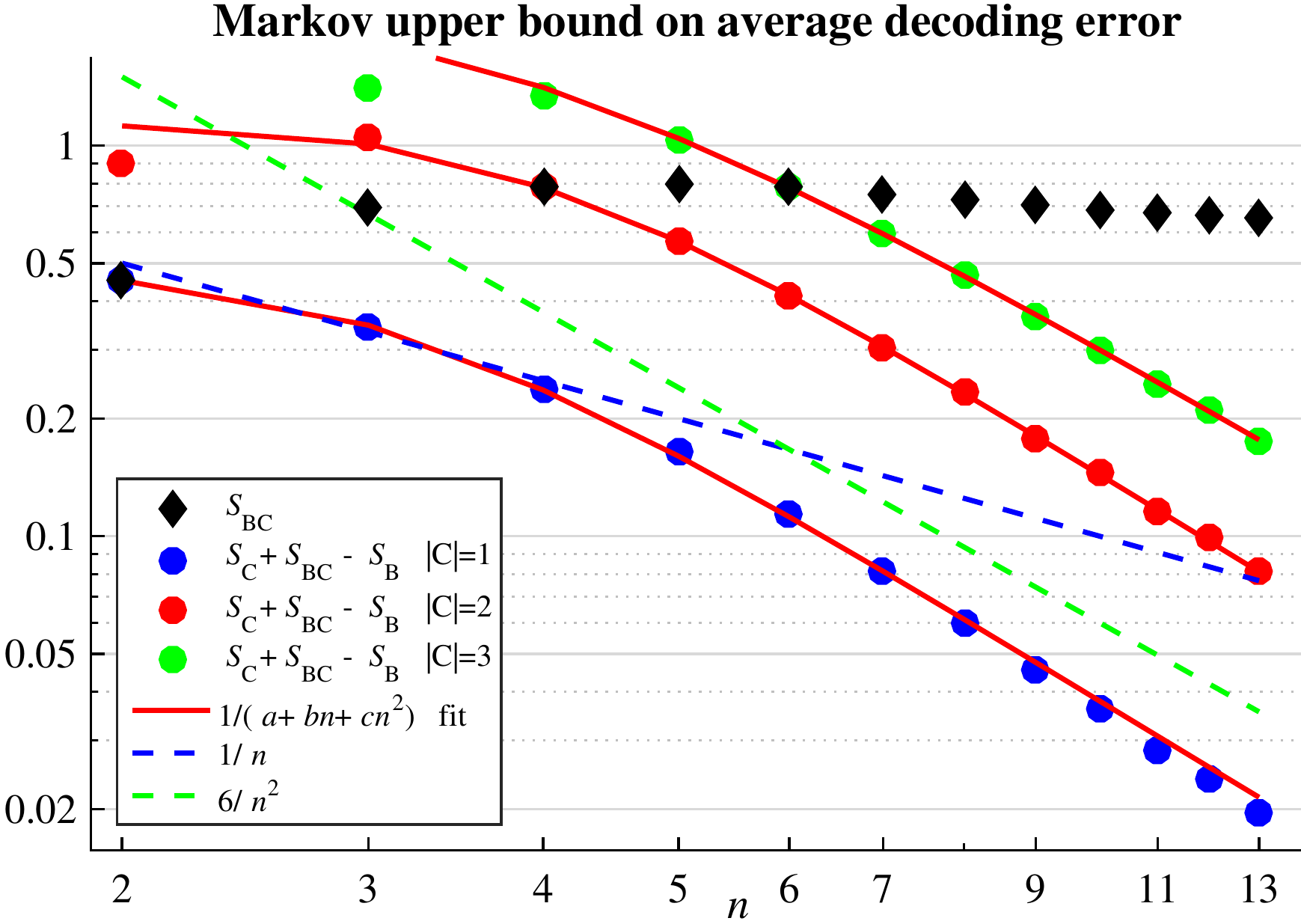}
		\caption{
			We consider $\rho_{BC}=\rho^{(\beta)}_n$ from eq.\ \eqref{eq:EvenThermal}.
			The number of physical qubits is given by $n=|BC|$ and take $\beta = 0.2n$ to maintain the global entropy $S_{BC}$ (black diamonds) approximatively constant.
			We plot CMI of eq.\  \eqref{eq:MarkovQuantitySimplified} associated to a small sub-systems $C$ using blue, red and green dots, corresponding to $|C|=1,2,3$.
			These provide convincing numerical evidence that the CMI converges quadratically for large $n$.
			We include $1/n$ (dashed blue) and $6/n^2$ (dashed green) lines as visual guides to the eye as well as Lorenzian fits to the data (solid red); we omit small $n < 2|C|$ from fitting.
		}\label{fig:MarkovianityConvergence}
	\end{center}
\end{figure}

Take $C$ to be a contiguous subregion of the lattice ($C \subset [1,n]$).
While there are numerous analytical calculations of entanglement entropies in CFTs
\cite{Holzhey, Korepin, Calabrese2009,Calabrese2011, Hubeny2013},
we are unaware of one allowing the evaluation of eq.\  \eqref{eq:MarkovQuantitySimplified} in terms of the joint scaling $\beta \propto n \rightarrow \infty$ and for a specific parity sector to a precision which goes beyond constant terms (whereas cancellation is only confirmed for divergent terms).
We therefore numerically calculate the CMI in eq.\ \eqref{eq:MarkovQuantitySimplified} for $\rho^{(\beta)}_\text{even}$ and confirm that there is a non-trivial increase in protection of the encoded information as $n$ increases, see fig.~\ref{fig:MarkovianityConvergence}.

{\bf The holographic interpretation.}
We now turn to connecting insights from quantum source-channel coding with
holography.
Recently, a connection between the holographic approach to quantum gravity \cite{Susskind1995, Hooft1993, Maldacena1998, witten1998anti, Verlinde2011} and quantum information notions such as entanglement \cite{Ryu2006}, compression \cite{Czech2015} and QECCs \cite{Verlinde2013, Almheiri2015, Pastawski2015, Hayden2016, Harlow2016,Pastawski2017} has emerged.
Toy models discussed so far \cite{Pastawski2015,Harlow2016} build upon notions of subspace coding, where the code space corresponds to the 'low energy subspace'.
In contrast, source-channel codes allow interpreting the full holographic duality as a unitary isomorphism between bulk and boundary Hilbert spaces, while preserving error correcting properties  to a certain extent.

In the holographic dictionary, CFT thermal states are interpreted as being dual to static black holes, where the black hole area is given by the thermal entropy in Planck units~\cite{Maldacena2003, VanRaamsdonk2017};
Constant black hole area corresponds to constant thermal entropy.
It has furthermore been argued that certain purifications of the thermal state such as the thermofield double $\ket{\Psi_{TFD}}$ admit a geometric interpretation beyond the black hole horizon as wormholes ~\cite{Maldacena2013}.

\begin{figure}[t!]
	\begin{center}
		\includegraphics[width=.7\columnwidth]{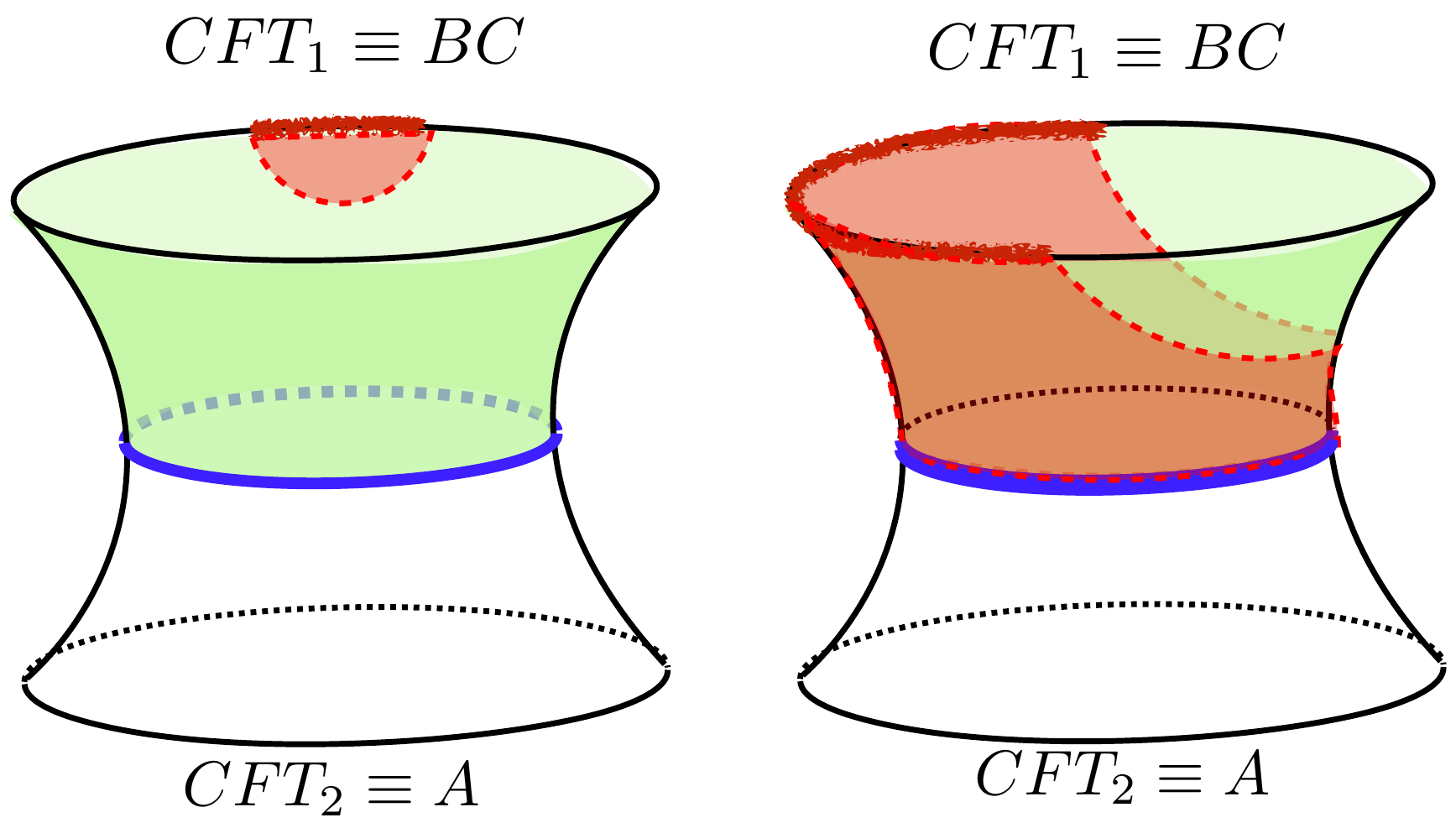}
		\caption{
			Illustration of the wormhole geometry, which corresponds to thermofield double state $\ket{\Psi_{TFD}}$.
			In our interpretation, only $\text{CFT}_1 \equiv BC$ is physical whereas $\text{CFT}_2 \equiv A$ is simply included for illustrative purpose.
			The bulk regions in green illustrates the \emph{entanglement wedge} of $B$ \cite{Hubeny2007, Czech2012}, this is the region contained between $B$ and the minimal bulk surface $\chi_B$ separating $B$ from its boundary complement $AC$.
			Left: A small CFT subregion $C$ is lost (red boundary) and the entanglement wedge of $B$ (green) reaches all the way up to the wormhole throat or black hole horizon (blue).
			This corresponds to accurate reconstruction on $B$ of thermally active degrees of freedom.
			Right: The CFT region $C$ that has been lost is large (roughly half) and the entanglement wedge of $B$ (green) no longer touches the wormhole throat.
			In this scenario, it is impossible to recover any of the information associated to the thermally active degrees of freedom.
			See Ref.\ \cite{Pastawski2017} for a more in depth development of the code theoretic interpretation of holographic geometries.
		}\label{fig:HolographicPicture}
	\end{center}
\end{figure}

Recently, a code theoretic interpretation for general holographic geometries has been put forth~\cite{Pastawski2017}.
In this interpretation, black hole horizons are interpreted as logical degrees of freedom carrying an amount of information equal to their horizon area.
Corresponding CFT degrees of freedom are interpreted as physical degrees of freedom.
In this setting, the degrees of freedom associated to a black hole are recoverable from a region $B$ of the boundary CFT if and only if the minimal surface $\chi_B$ separating $B$ from its boundary complement includes the black hole horizon.
Using the Ryu-Takayanagi prescription \cite{Ryu2006} of identifying the area of minimal bulk surfaces with the entropy of corresponding boundary regions (see Fig. \ref{fig:HolographicPicture}) the geometric condition can be identified as the CMI condition \eqref{eq:MarkovQuantitySimplified}.
The critical Ising CFT is not expected to be dual to a classical geometry, since the AdS radius is the Planck scale ($N=O(1)$) and the central charge $c=1/2$ of the model are of order one \cite{Castro2012};
Nevertheless, our numerical findings are in qualitative agreement with an interpretation of a Schwartzschild-AdS geometry (See Fig. \ref{fig:HolographicPicture}).

{\bf Conclusion.} We have shown that the thermal state of a CFT can be interpreted as a quantum source-channel code which is approximately protected from small local errors.
While we provided strong numerical evidence that the relevant CMI goes to zero as $n^{-2}$,
a rigorous analytic treatment of the CMI convergence is still lacking.
One obstruction for such an analytic treatment is the crucial projection onto a super-selection sector, which spoils the non-interacting character of the model.
However, it would be interesting to see if the recent analytical results of Ref.\ \cite{Lashkari2016} can be adapted to the interpretation of CFTs as source-channel codes.

It is clear that as $|C|/n$ increases, our lower bound on the recovery fidelity suffers.
Future work should elucidate whether the precise scaling predicted by AdS/CFT entanglement wedge hypothesis \cite{Hubeny2007} for the transition between recoverability and scrambling is reproduced by simple CFT models in terms of $n$, $S_{BC}$ and $|C|$.
While we have focused on global recovery due to the small accessible system size,
it is natural to considering local recovery, wherein the recovery map is restricted to act on a small patch of the physical system (i.e. $C$ plus some constant buffer region around it) as is done in recent work on approximate quantum error correction \cite{Flammia2016,Kim2016, Cotler2017}.
In this case, geometric considerations \cite{Pastawski2017, Kim2016} indicate that the necessary buffer region around $C$ will have a width proportional to the size of $C$ instead of constant.
Our example suggests that restricting to a specific topological sector of the CFT might be necessary in order to allow non-trivial reconstruction of the thermally active degrees of freedom -- indicating that gauge invariance may play a role in the QECC features as suggested by Refs.~\cite{Mintun2015, Freivogel2016}.
It is the hope that the present work can significantly stimulate such further endeavors
identifying the precise connection between holography and quantum error correction.

\begin{acknowledgments}
\textit{Acknowledgements.} FP would like to acknowledge David Aasen and emphatically Isaac Kim and Michael J. Kastoryano for fruitful discussions and comments.
We thank an anonymous TQC referee for pointing out Ref. \cite{Barnum2002}.
FP gratefully acknowledges funding provided by the FUB through ERC project (TAQ) as well as the AvH foundation, JE by the ERC (TAQ), the
DFG (CRC 183, EI 519/7-1), the EC (AQuS), and the Templeton Foundation.
This research was supported in part by the National Science Foundation under Grant No.\ NSF PHY11-25915.
HW acknowledges funding through ERC (TAQ) and the Studienstiftung des Deutschen Volkes.
\end{acknowledgments}


\begin{thebibliography}{10}
\bibitem{Trivial}
	{We say the $1/n$ scaling is trivial because it can be achieved by recording a
		single copy of the logical information locally on one physical site and doing
		so uniformly (coherently or not) over the $n$ physical sites. Lossing a site
		then corresponds to losing a fraction $1/n$ of the information.}
\end{thebibliography}

\newpage

\appendix

\myappendix{
	
\section{Appendix A: Comparison to subspace coding}\label{sec:SubspaceCoding}
In this appendix, we discuss how our framework incorporates the case of traditional QECCs. In QECCs, the logical information of a $D$-dimensional quantum system is encoded isometrically into a $D$-dimensional code-subspace  $\cC$ of a high-dimensional physical Hilbert-space $\mathcal{H}$.
All pure states in this sub-space are considered equally likely.
In our set-up this corresponds to choosing as probability measure $\mu$ the \emph{Haar measure} $\mu_{\mathrm{Haar}}$ over $\cC$.
This is the unique measure which is fully invariant under unitary operators within $\cC$ and represents a flat prior within $\cC$.
The density matrix corresponding to such an ensemble is given by the normalized projector on the code-subspace, i.e., a density matrix with low rank on the full Hilbert-space $\mathcal{H}$.

In such a set-up its customary to drop $\mu$ from the average fidelity and entanglement fidelity, and simply write $\overline{F}(\cE)$ and $F_e(\cE)$, respectively. A standard benchmark for the recovery of information is then given by the average fidelity of the code
\begin{align}
\overline{F}(\cE) = \int_{\psi} \langle \psi
|(\mathcal{R} \circ \mathcal{N}) ( | \psi \rangle \langle \psi |)\!\mid \!\psi\rangle  \mu_{\mathrm{Haar}}(\psi)
 d \psi.
\end{align}
Thus, our approach incorporates standard subspace coding in a natural way.
The following simple relation between average fidelity $\bar{F}$ and entanglement fidelity $F_e$ of a channel $\cE$ has been proven in Ref.\ \cite{Horodecki1999} (see also Ref.\ \cite{Nielsen2002})
\begin{align}
\bar{F}(\cE) = \frac{D F_e(\cE) + 1}{D +1}.
\label{eq:AvgVsEntanglementFidelity}
\end{align}
The RHS of eq.\  \eqref{eq:AvgVsEntanglementFidelity} can be further simplified  to $ \bar{F}(\cE) = F_e(\cE) + ({1 - F_e(\cE)})/({D +1})$ where it becomes explicit that for large code-space dimension $D=|\cC|$, the average fidelity and the entanglement fidelity are essentially equivalent.

\section{Appendix B: Extremality of Gaussian fermionic states}\label{sec:GaussianStates}

In this appendix, we show that fermionic Gaussian states extremize the quantity in eq.\ \eqref{eq:MarkovQuantitySimplified} in the following sense.
To every fermionic state $\rho$ we can associate a \emph{covariance matrix}, which collects all the second moments of the Majorana operators in this state.
 These second moments uniquely determine a Gaussian state $\omega$ with the same second moments.
 The quantity in eq.\ \eqref{eq:MarkovQuantitySimplified} evaluated on this Gaussian state can only be larger than that of the original state $\rho$.


This statement can be proven as follows for general
pairs of quantum states $\rho$ and the Gaussian state $\omega$ with the same
second moments of Majorana operators as $\rho$.
To start with,
\begin{equation}
S(\rho_B)\leq
S(\omega_B),
\end{equation}
exploiting the well-known extremality of Gaussian states for
the von-Neumann entropy. Then,
\begin{equation}
S(\rho_{BC}\|\omega_{BC}) \geq
S(\rho_{C}\|\omega_{C}),
\end{equation}
from the contraction property of the quantum
relative entropy under completely positive maps. This gives
\begin{equation}
-S(\rho_{BC}) - {\rm tr}(\rho_{BC} \log\omega_{BC}) \geq -S(\rho_C) - {\rm
	tr}(\rho_C\log\omega_C).
\end{equation}
The Gaussian state is (a limit of) a matrix exponential of a
Hamiltonian quadratic in fermionic Majorana operators, implying $\Tr(\rho_{BC}\log \omega_{BC})=\Tr(\omega_{BC}\log \omega_{BC})$ and hence
\begin{equation}
-S(\rho_{BC}) - {\rm
	tr}(\omega_{BC}\log\omega_{BC}) \geq -S(\rho_C) - {\rm
	tr}(\omega_C\log\omega_C).
\end{equation}
Upon reordering the terms, it follows that
\begin{eqnarray}
S_C +
S_{BC} -S_B &=& S(\rho_C) + S(\rho_{BC}) - S(\rho_B)  \nonumber\\
&\leq &S(\omega_C)
+
S(\omega_{BC}) - S(\omega_B),
\end{eqnarray}
from which the assertion
follows.

\section{Appendix C: Encoding more information}\label{sec:Largek}
\begin{figure}[t]
	\begin{center}
		\includegraphics[width=1\columnwidth]{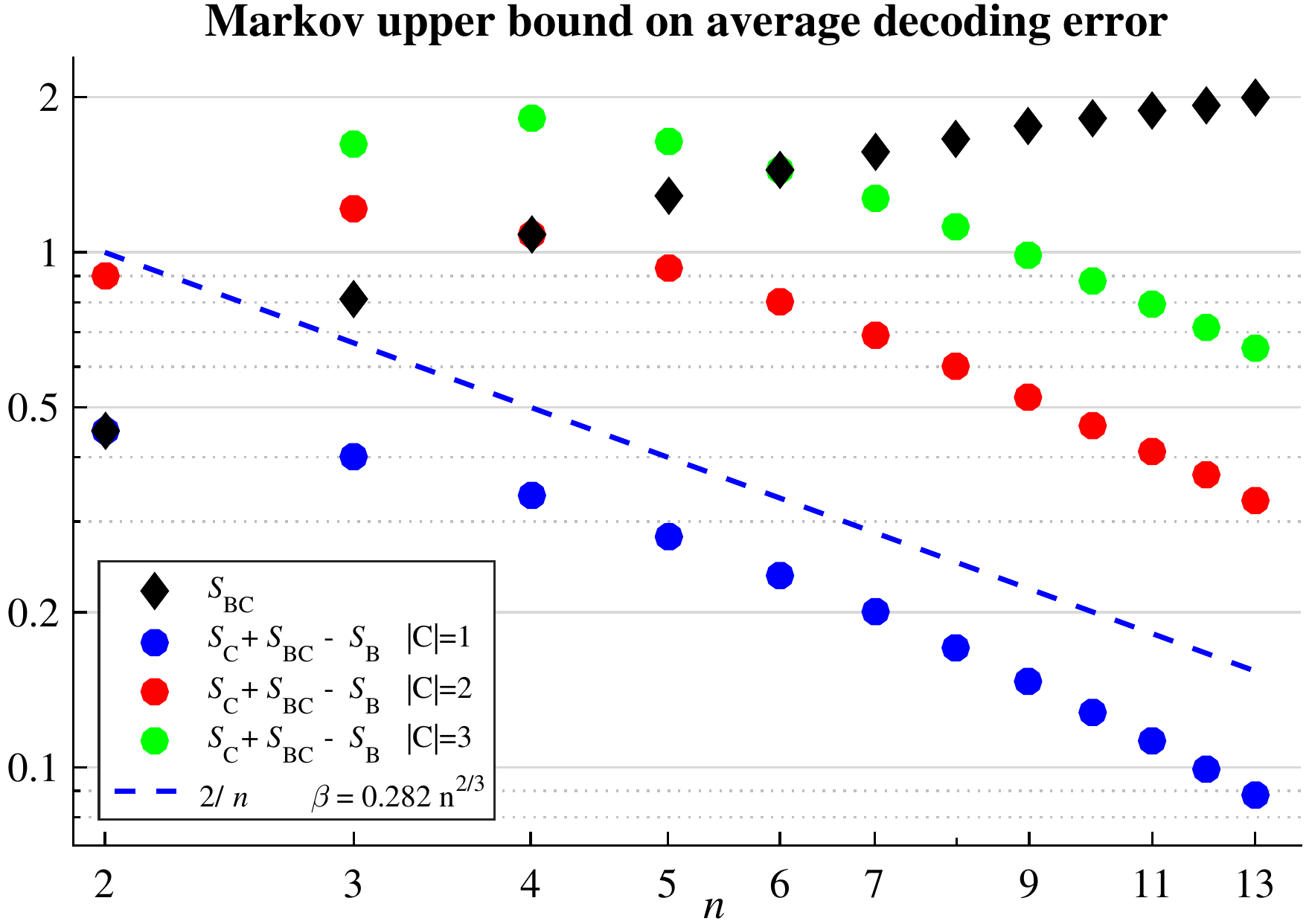}
		\includegraphics[width=1\columnwidth]{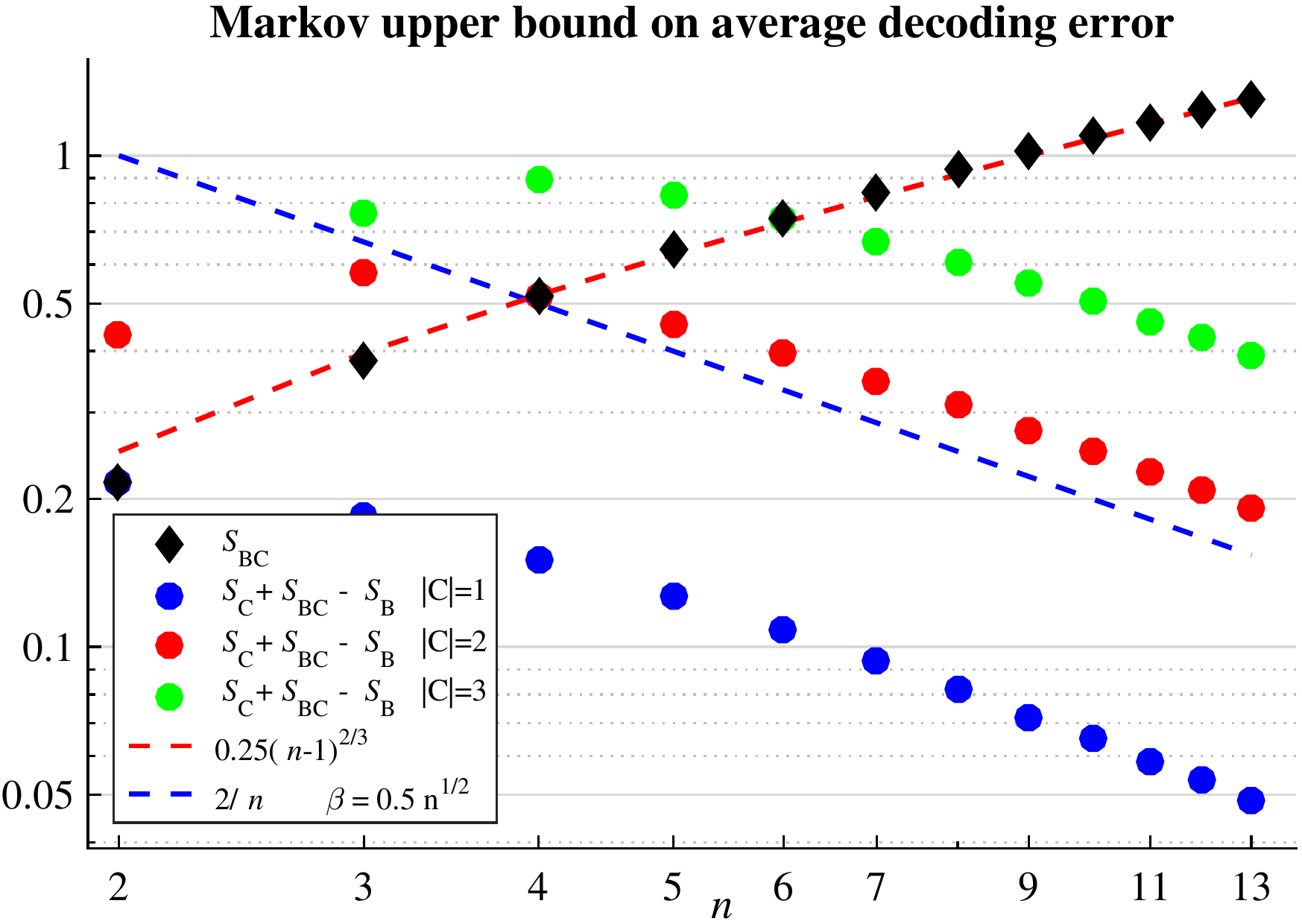}
		\caption{
			The top plot considers $\beta = 0.282n^{2/3}$ whereas the bottom plot considers $\beta = 0.5n^{1/2}$.
			In these figures we depict how the inclusion of a growing number of thermally active degrees of freedom impacts the recoverability from local noise.
			Although the Markov quantity of eq.\  \eqref{eq:MarkovQuantitySimplified} continues to converge, it does so more slowly than in the original case of constant thermal entropy.
			Namely, the slower $\beta$ increases, the faster $S_{BC}$ increases and the slower $S_C+S_{BC}-S_B$ converges to zero.		}\label{fig:MoreLogicals}
	\end{center}
\end{figure}
We may use the same Hamiltonian $H_{TF}$ as in the example to attempt to encode more logical information.
While it will be necessary to send $\beta \rightarrow \infty$ in order to get convergence of the Markov quantity, we may nevertheless do so more slowly than $\beta \propto n$.
To study this, we take $\beta$ scaling sublinearly as $n^{\alpha}$ with $n$ (with $0<\alpha <1$). 
The numerical results can be found in fig.\ \ref{fig:MoreLogicals}. We obtain a polynomial increase of thermal entropy $S_{BC}$ with $n$. 
While the numerics suggest the CMI of \eqref{eq:MarkovQuantitySimplified} continues to converge to zero for large $n$, it does so more slowly than the $n^{-2}$ convergence obtained for $\beta \propto n$.

\clearpage

}

\end{document}